\documentclass[%
 reprint,
amsmath,amssymb,
aps,
]{revtex4-2}
\usepackage{arydshln}
\usepackage{multirow}
\usepackage{ulem}
\usepackage{graphicx}% Include figure files

\usepackage{dcolumn}% Align table columns on decimal point
\usepackage{bm}% bold math

\usepackage[colorlinks,linkcolor=blue,citecolor=blue,urlcolor=blue]{hyperref}

\usepackage{booktabs}

\usepackage{mhchem}

\begin{document}

\preprint{}

\title{Multiband transport hierarchy and large Nernst effect in EuAuBi: Establishing a Nernst scaling for asymmetric multiband systems }

\author{Weimin Quan$^{1}$, Xiaodong Guo$^{1}$, Lingxiao Zhao$^{2,3}$, Shengwei Chi$^{1}$, Gang Xu$^{1}$, Zengwei Zhu$^{1,*}$ and Xiaokang Li$^{1,*}$}

\affiliation{
(1) Wuhan National High Magnetic Field Center, School of Physics, Huazhong University of Science and Technology, 430074 Wuhan, China\\ 
(2) Quantum Science Center of Guangdong-HongKong-Macao Greater Bay Area, Shenzhen 523335, China \\
(3) Department of Physics, Southern University of Science and Technology, Shenzhen, China 
}

\date{\today}

\begin{abstract}
In correlated materials, coexisting pockets of vastly different carrier densities raise two fundamental questions: which pocket governs the various transport coefficients, and does the conventional Nernst scaling $\nu/T \propto \mu/E_F$, originally derived for single-band systems, still hold? We address both questions in the polar semimetal EuAuBi, where a dilute electron pocket ($n_e \sim 10^{16}~\mathrm{cm}^{-3}$) coexists with a dense hole pocket ($n_h \sim 10^{21}~\mathrm{cm}^{-3}$). We find a clear hierarchy: the hole pocket dominates the longitudinal resistivity; the Hall effect crosses from electron- to hole-dominance with increasing field; the Seebeck coefficient is dominated by the electron pocket at low temperature and by both pockets at high temperature. Remarkably, the Nernst effect is governed entirely by the ultrahigh-mobility electron pocket, yielding a large low-field signal of $\sim 5~\mu\mathrm{V/K}$ near 1~T at 202~K, comparable to anomalous Nernst signals in magnetic Weyl semimetals. By analyzing the two-band thermoelectric conductivity, we show that the Nernst coefficient follows a scaling $\nu/T \propto \mu_e/{E_{F, tot}}$. This scaling originates from a compensation between the electron-to-hole conductivity ratio and the Fermi-energy ratio, establishing that the large Nernst effect is a semiclassical multiband phenomenon rather than a topological Berry-curvature contribution. This understanding advances the thermoelectric transport physics of multiband electronic systems and offers a guiding principle for low-field transverse thermoelectric design.

\end{abstract}

\maketitle

Complex Fermi surfaces often host multiple pockets with carrier densities differing by orders of magnitude \cite{ZhuWTe2,PRXCd3As2,matusiak2017thermoelectric,twoband,PRBTwoband,cd3as2largenernst}. Electrical transport is typically dominated by the most conductive pocket, while thermoelectric coefficients are far more sensitive to low-density, high-entropy pockets \cite{behnia2015fundamentals,PhysRevLett.98.076603,behnia2022measured}. Multiband effects have been studied in compensated semimetals \cite{ali2014large,li2022colossal,1948theory}, but a comprehensive investigation in systems where an extremely small pocket coexists with a much larger one remains largely unexplored. In such asymmetric systems, a pressing open question is whether the conventional Nernst scaling $\nu/T \propto \mu/E_F$ \cite{behnia2009nernst,behnia2016nernst}, derived rigorously for single-band systems, still holds, or whether it requires fundamental revision. Resolving this question---and establishing the corresponding transport hierarchy---is the central motivation of the present study.

\begin{figure*}[ht]
\centering
\includegraphics[width=0.95\linewidth]{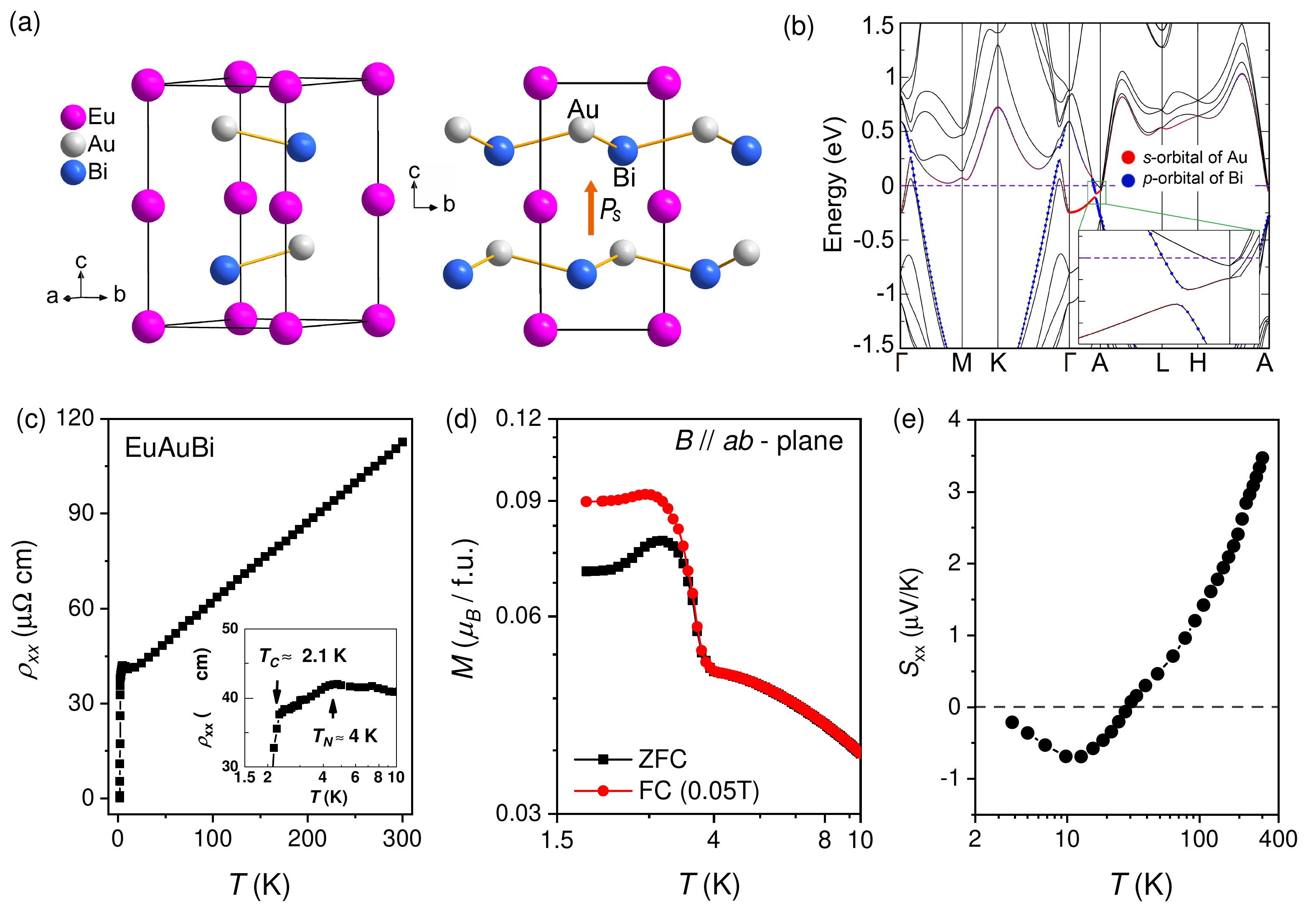} 
\caption{\textbf{Longitudinal transport in EuAuBi.}
(a) Crystal structure of EuAuBi, consisting of a layered Au–Bi honeycomb lattice that gives rise to a polar crystal structure along the \(c\) axis. 
(b) Calculated band structure of EuAuBi. Red and blue symbols denote the projections onto Au-\(s\) and Bi-\(p\) orbitals, respectively. Near the A point, band inversion produces a band crossing below the Fermi level, giving rise to a small electron pocket, as highlighted in the inset \cite{SM}.
(c) Temperature dependence of the resistivity \(\rho_{xx}\) from 1.8 to 300 K. The antiferromagnetic transition and superconducting transition are marked by black arrows in the inset, respectively.
(d) Temperature-dependent magnetization from 1.8 to 10 K. Around 4 K, the FC and ZFC curves start to diverge, indicating the formation of an antiferromagnetic order.
(e) Temperature dependence of the Seebeck coefficient \(S_{xx}\) in EuAuBi. %The temperature dependence of ${S_{xx}}/T$ is shown in the inset, indicating the Fermi energy is $\sim$ 0.67 eV.
}
\label{1}
\end{figure*}

EuAuBi is a polar semimetal with a layered Au-Bi honeycomb lattice and strong spin–orbit coupling \cite{EuAuBiJPS,EuAuBiPRB}. It enters an antiferromagnetic state at $\sim 4$~K and becomes superconducting below $2.1$~K \cite{EuAuBiJPS}. Crucially, band-structure calculations reveal a small electron pocket with very low carrier density near the A point, coexisting with a much larger hole pocket \cite{Xu,PRLLiGaGe,Rashba}. This extreme density mismatch, together with the material's rich phase diagram \cite{EuAuBiPRB}, makes EuAuBi an ideal platform to disentangle multiband transport hierarchy and test the validity of the conventional Nernst scaling.

In this work, we disentangle the multiband transport hierarchy in EuAuBi via resistivity, Hall, Seebeck, and Nernst measurements. We reveal a clear division of roles. (i) the hole pocket dominates longitudinal resistivity; (ii) the Seebeck coefficient is governed by the dilute electron pocket at low temperatures and by both pockets at high temperatures; and (iii) the Nernst effect is dictated solely by the ultrahigh-mobility electron pocket, yielding a large low-field signal of $\sim5~\mu\mathrm{V/K}$ near 1~T at 202~K, comparable to anomalous Nernst responses in magnetic Weyl semimetals \cite{Diracdispersion,ding2019,sakai2018giant}. By formulating the Nernst response in terms of the thermoelectric conductivity $\alpha_{xy}$, we demonstrate that the Nernst coefficient follows a scaling $\nu/T \propto \mu_e/E_{F, tot}$ that arises from a compensation between the electron-to-hole conductivity ratio and the Fermi energy ratio, revealing that the enhanced Nernst effect originates from the semiclassical transport of the high-mobility electron pocket (high mobility and large entropy) rather than from topological Berry-curvature contributions \cite{Prb-weyl-dirac,Diracdispersion,BERRY}.
Our findings reveal a clear hierarchy of transport responses. This insight not only advances the fundamental understanding of transport in multiband systems but also provides a guiding principle for transverse thermoelectric applications at low magnetic fields. 
%in which distinct pockets dominate different transport channels, and establish that the Nernst effect can be governed by a dilute, high-mobility pocket through conventional mechanisms. First, the longitudinal resistivity is dominated by the large hole pocket at all temperatures, and the Hall effect evolves from electron-dominance at low fields to hole-dominance at high fields. Second, for thermoelectric transport, the Seebeck coefficient is governed by the dilute electron pocket at low temperatures and by both pockets jointly at high temperatures. Finally, the Nernst effect is dictated entirely by the tiny, ultrahigh-mobility electron pocket across the whole temperature range, yielding an unusually large low-field Nernst signal of $\sim 5~\mu\mathrm{V/K}$ near 1~T at 202~K, comparable to the anomalous Nernst response of magnetic Weyl semimetals \cite{Diracdispersion,ding2019,sakai2018giant}.

\begin{figure*}[ht]
\centering
\includegraphics[width=0.95\linewidth]{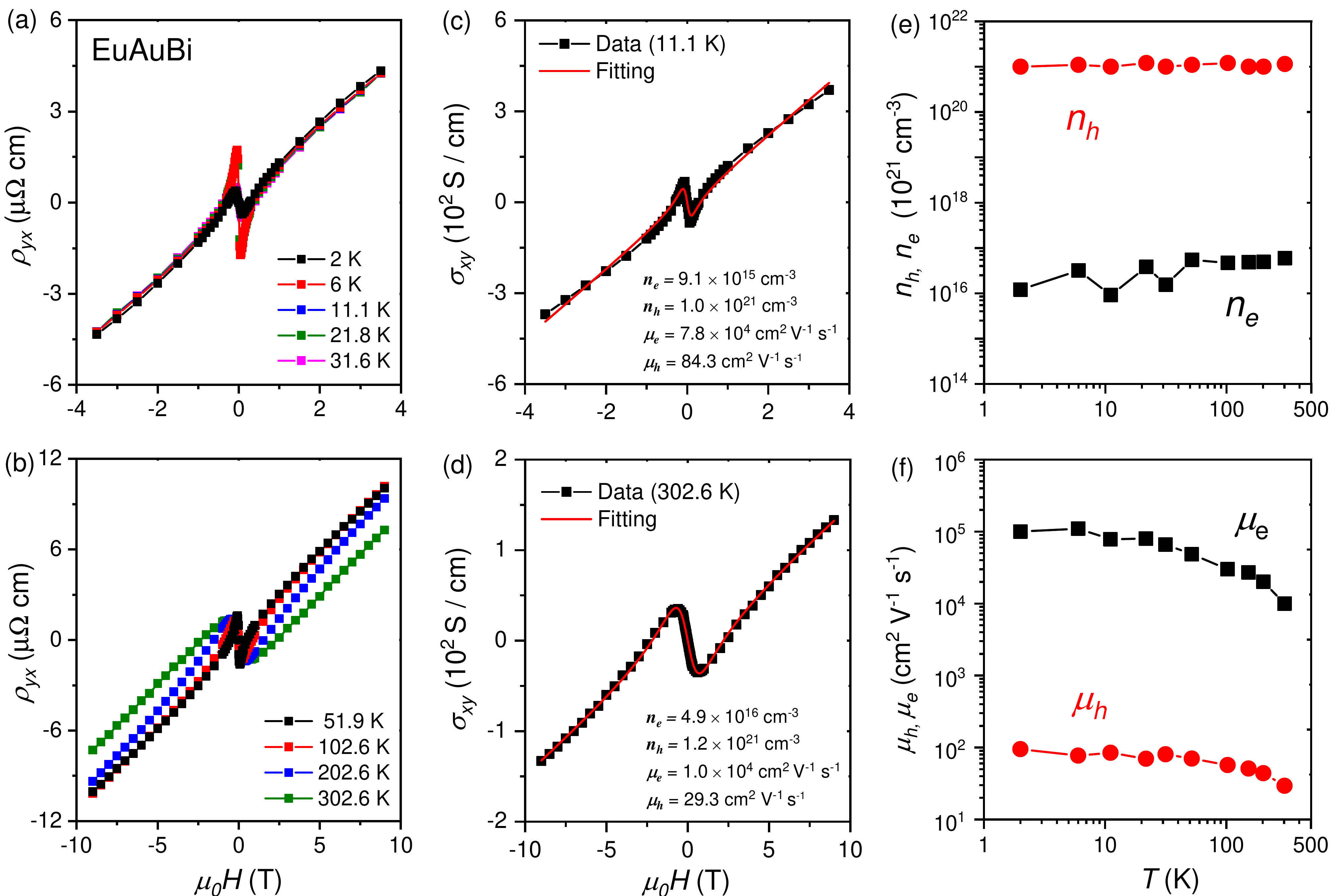} 
\caption{\textbf{Multicarrier Hall transport in EuAuBi.}
\textbf{(a, b)} Magnetic-field dependence of the Hall resistivity $\rho_{yx}$ at selected temperatures. A pronounced step-like feature develops in the low-field region and becomes increasingly prominent with increasing temperature, indicative of multicarrier transport. 
\textbf{(c, d)} Hall conductivity $\sigma_{xy}$ at 11.1 K and 302.6K, converted from $\rho_{yx}$, together with the corresponding fit to the two-band model. The fitting reveals two distinct carrier components: a large hole pocket with high carrier density and low mobility, and a small electron pocket with low carrier density but high mobility.
\textbf{(e)} Temperature dependences of the carrier densities extracted from the two-band analysis, where $n_h$ and $n_e$ denote the hole and electron densities, respectively. \textbf{(f)} Temperature dependences of mobilities. Here, $\mu_h$ and $\mu_e$ represent hole and electron mobilities, respectively.
}
\label{2}
\end{figure*}

\begin{figure*}[ht]
\centering
\includegraphics[width=1\linewidth]{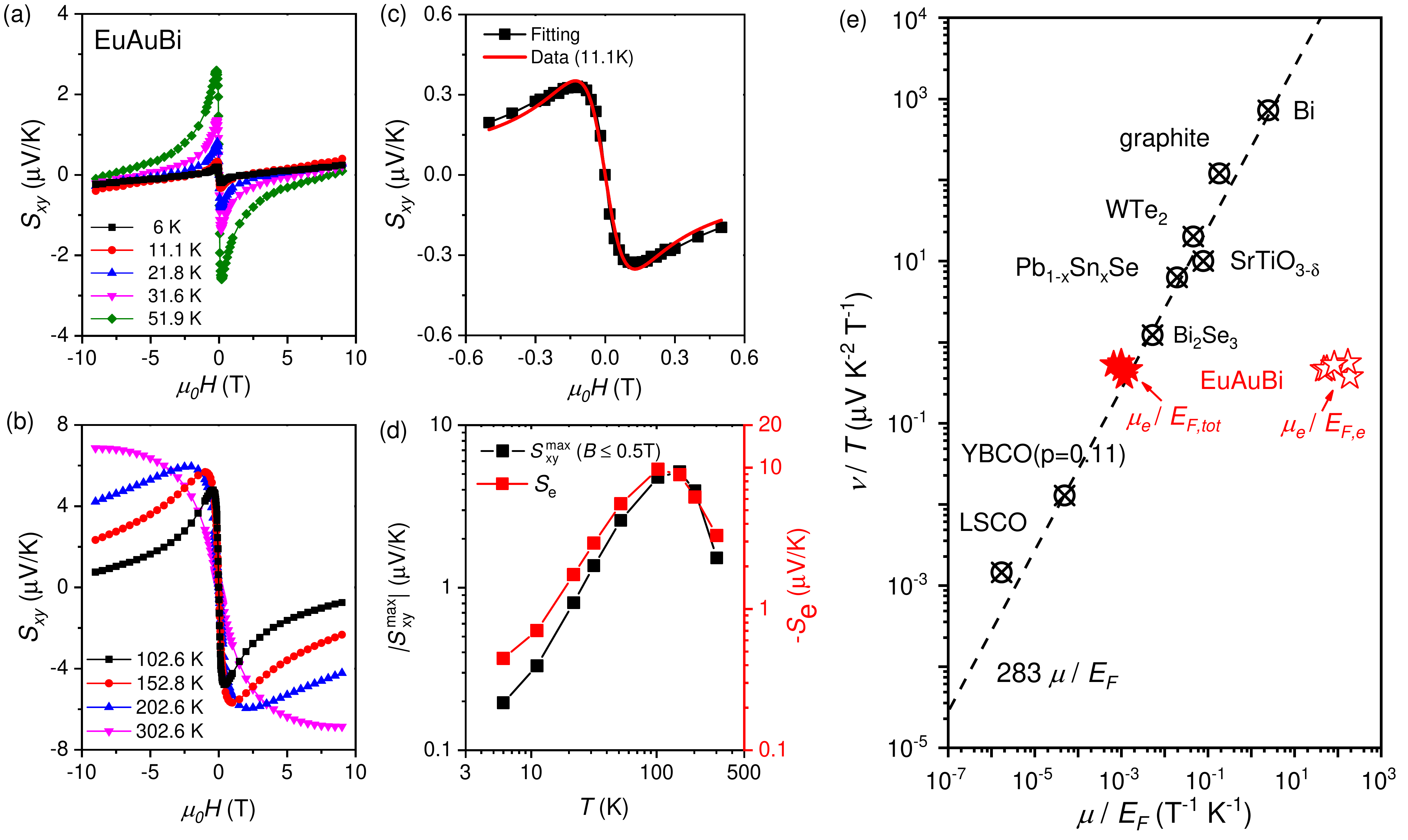} 
\caption{\textbf{Large Nernst effect in EuAuBi.}
\textbf{(a, b)} Magnetic-field dependence of the Nernst signal for EuAuBi.
\textbf{(c)} Nernst signal $S_{xy}$ at 11.1 K, together with the corresponding fit to the single-band thermoelectric model (see text). In the fitting, the carrier mobilities were fixed to the values extracted from the two-band Hall analysis.
\textbf{(d)} The maximum Nernst signal value $S_{xy}^{max}$ and fitting amplitude -$S_e$ at each temperature point. 
\textbf{(e)} Nernst coefficient divided by temperature, \(\nu/T\), plotted as a function of \(\mu/E_F\) in different materials. The dashed line represents the theoretical scaling relation \(\nu/T \propto \mu/E_F\) \cite{behnia2016nernst}. For EuAuBi (red stars), the scaling is examined using two choices of Fermi-energy scale: the electron-pocket Fermi energy $E_{F,e}$ and the effective total Fermi-energy scale $E_{F,tot}$. The data plotted with $E_{F,e}$ deviate from the dashed line, whereas those plotted with $E_{F,tot}$ collapse onto the expected scaling.
}
\label{3}
\end{figure*}

EuAuBi crystallizes in the hexagonal space group $P6_{3}/mc$, characterized by a layered Au-Bi honeycomb lattice that intrinsically breaks spatial inversion symmetry and yields a polar crystal structure, as shown in Fig.~\ref{1}\textbf{a} \cite{EuAuBiJPS,Xu}. Due to the strong spin–orbit coupling of the heavy Bi and Au elements, this crystal provides a natural platform to realize topological band states \cite{PRLLiGaGe,Rashba}. Fig.~\ref{1}\textbf{b} shows the band calculations for the antiferromagnetic phase. Around the A point, band inversion gives rise to a band crossing feature below the Fermi level, resulting in a small electron pocket, as highlighted in the inset \cite{Xu,PRLLiGaGe,Rashba}. As discussed below, the coexistence of this small electron pocket with a much larger conventional hole pocket leads to a clear hierarchy in the transport response.

Fig.~\ref{1}\textbf{c} shows the temperature dependence of the longitudinal resistivity $\rho_{xx}$. Reflecting the metallic nature of EuAuBi, $\rho_{xx}$ is relatively small, reaching 110~$\mu\Omega\,\mathrm{cm}$ at room temperature and decreasing nearly linearly upon cooling to 10~K. An upturn appears below this temperature, accompanying the onset of antiferromagnetic order, as confirmed by the separation of the field-cooled and zero-field-cooled magnetization near 4 K (Fig.~\ref{1}\textbf{d}). At lower temperature, EuAuBi becomes superconducting below $T_c = 2.1$~K, in agreement with previous reports \cite{EuAuBiJPS}. Within a multiband picture, the zero-field longitudinal conductivity is given by $\sigma_{xx}=e(n_h\mu_h~+~n_e\mu_e)$, where $n_h$ ($n_e$) and $\mu_h$ ($\mu_e$) denote the hole (electron) density and mobility, respectively. As discussed below, the Hall derived parameters give $n_h\mu_h \sim 10^{22}$ and $n_e\mu_e \sim 10^{20}~\mathrm{cm^{-1}V^{-1}s^{-1}}$, indicating that $\sigma_{xx}$ is overwhelmingly dominated by the dense hole pocket.

Fig.~\ref{1}\textbf{e} displays the temperature dependence of the Seebeck coefficient $S_{xx}$, which is nonmonotonic, showing a broad negative minimum near 10 K, a sign change around 30 K, and positive values at higher temperatures. Within a two-band model, the total Seebeck coefficient is given by the conductivity-weighted average $S = (\sigma_e S_e^0 + \sigma_h S_h^0)/(\sigma_e + \sigma_h)$. At low temperatures, the electron pocket, owing to its small Fermi energy (Fig. \ref{1}\textbf{b}) and the large magnitude of its negative diffusive thermopower $S_e^0$, dominates the total $S_{xx}$. As temperature increases, the positive hole contribution $S_h^0$ becomes increasingly important, eventually driving the sign change of $S_{xx}$. This evolution demonstrates that the Seebeck response is governed primarily by the dilute electron pocket at low temperatures and by both pockets at higher temperatures, revealing a clear multiband hierarchy in the thermoelectric response. %The inset shows that $|S_{xx}|/T$ reaches 0.055 $\mu\mathrm{V/K^2}$ at 4 K, corresponding to an effective Fermi-energy scale of $E_F \simeq 0.67$~eV.

A more striking manifestation of the two-pocket coexistence appears in the Hall effect. Fig.~\ref{2}\textbf{a} and \textbf{b} show $\rho_{yx}(B)$ for EuAuBi at different temperatures. At high fields, $\rho_{yx}$ is nearly linear, while a pronounced low-field step-like feature develops with increasing temperature. This reflects a crossover in the Hall response: the high-mobility electron pocket dominates at low fields, producing a negative slope, whereas its contribution saturates at higher fields and the nearly linear hole contribution takes over. To quantify this behavior, we converted $\rho_{yx}$ to $\sigma_{yx}=-\rho_{yx}/(\rho_{xx}^2+\rho_{yx}^2)$ and fitted it using the semiclassical two-band expression:
\begin{equation}
  \sigma_{yx}(B) = \frac{n_e e \mu_e^2 B}{1 + (\mu_e B)^2} - \frac{n_h e \mu_h^2 B}{1 + (\mu_h B)^2}.
  \label{eq:two band}
\end{equation}

An example of the two-band fit at 11.1 K is shown in Fig.~\ref{2}\textbf{c}. The fitting resolves two highly distinct carrier components: a large hole pocket with a high carrier density of $n_h \sim 1\times10^{21}$~cm$^{-3}$ and a low mobility of $\mu_h \sim 84.3$~cm$^{2}$V$^{-1}$s$^{-1}$, and a tiny electron pocket with a much lower density of $n_e \sim 9.1\times10^{15}$~cm$^{-3}$ but an exceptionally high mobility on the order of $10^{4}$~cm$^{2}$V$^{-1}$s$^{-1}$. A similar hierarchy is observed at 302.6 K, as shown in Fig.~\ref{2}\textbf{d}. The temperature dependences of the extracted carrier densities and mobilities are summarized in Fig.~\ref{2}\textbf{e} and \textbf{f}, respectively. Over the measured temperature range, the electron density remains nearly five orders of magnitude smaller than the hole density, whereas the electron mobility is about three orders of magnitude larger than the hole mobility. This pronounced density–mobility contrast provides the basis for the hierarchy of transverse transport. As discussed below, the distinct roles of the two pockets are most clearly manifested in the Nernst effect.

%Within a multiband picture, the zero-field conductivity satisfies $\sigma = e\,(n_h\mu_h + n_e\mu_e)$. From the Hall-derived parameters, we find $n_h\mu_h \approx 3\times10^{22}$~cm$^{-1}$V$^{-1}$s$^{-1}$ and $n_e\mu_e \approx 2\times10^{20}$~cm$^{-1}$V$^{-1}$s$^{-1}$, indicating that the longitudinal conductivity is overwhelmingly dominated by the dense hole pocket, while the dilute ultrahigh-mobility electron pocket makes only a negligible contribution in zero field. 

Fig.~\ref{3}\textbf{a} and \textbf{b} show the adiabatic Nernst thermopower $S_{xy}$ as a function of magnetic field. In contrast to the conventional linear Nernst response of single-band metals, $S_{xy}$ exhibits a pronounced low-field peak. At 202.6 K, the peak reaches $\sim 5~\mu$V/K near 1 T, comparable in magnitude to the anomalous Nernst signals reported in magnetic Weyl semimetals such as Co$_3$Sn$_2$S$_2$ ($3~\mu$V/K) \cite{ding2019} and Co$_2$MnGa ($6~\mu$V/K) \cite{sakai2018giant}.    

To capture the multiband origin of this response, we analyze the thermoelectric conductivity tensor \(\alpha_{ij}\), which, like the electrical conductivity tensor \(\sigma_{ij}\), is additive over independent pockets. In the semiclassical two-band picture, the transverse thermoelectric conductivity can be written as
\begin{equation}
  \alpha_{xy}(B)=\alpha_e^0\frac{\mu_e B}{1+(\mu_e B)^2}
  +\alpha_h^0\frac{\mu_h B}{1+(\mu_h B)^2},
  \label{eq:alphaxy}
\end{equation}
where \(\alpha_{e(h)}^0\) are fitting amplitudes associated with the electron and hole pockets. In the low-field regime, where the Hall angle is small and the longitudinal conductivity is dominated by the hole pocket (\(\sigma_{xx}\approx \sigma_h\)), this expression reduces approximately to (Supplementary Materials Note 2 \cite{SM})
\begin{equation}
  S_{xy}(B)\approx S_e\frac{\mu_e B}{1+(\mu_e B)^2}
  +S_h\frac{\mu_h B}{1+(\mu_h B)^2},
  \label{eq:Sxy}
\end{equation}
where \(S_{e(h)}\) are effective fitting amplitudes with $S_{e(h)} = \alpha_{e(h)}^0 / \sigma_h$. Because $\mu_e$ exceeds $\mu_h$ by nearly three orders of magnitude, the electron term in Eq.~(\ref{eq:Sxy}) dominates the low-field response, giving rise to the steep initial increase and the subsequent peak when $\mu_e B \sim 1$. The hole term remains essentially linear in the measured field range and is far too small to compensate the decay of the electron contribution, so the overall field profile is entirely shaped by the electron pocket. 

To verify this interpretation, we fitted the low-field Nernst data at 11.1~K with the single-component form $S_{xy}=\frac{S_e \mu_e B}{1+(\mu_e B)^2}$, where $\mu_e$ was fixed to the value obtained from the Hall analysis. As shown in Fig.~\ref{3}\textbf{c}, the fit reproduces the data well and yields $S_e=-0.7~\mu\mathrm{V/K}$, supporting the assignment of the low-field Nernst response to the tiny ultrahigh-mobility electron pocket (Supplementary Materials Note 3 \cite{SM}). Fig.~\ref{3}\textbf{d} shows that $|S_{xy}^{\max}|$ and $-S_e$ follow the same temperature dependence, supporting the assignment of the low-field Nernst peak to the effective thermopower of the electron pocket. 

In the low-field limit, Eq.~(\ref{eq:Sxy}) gives the Nernst coefficient as 
$\nu \simeq (S_e\mu_e+S_h\mu_h) \simeq S_e\mu_e$, because $\mu_e\gg\mu_h$. 
The magnitude of the effective electron thermopower scales as 
$S_e \simeq \frac{\alpha_e^0}{\sigma_h}
\simeq \frac{\pi^2}{3}\frac{k_B^2T}{e} \frac{\sigma_e}{\sigma_h}\frac{1}{E_{F,e}}$, 
where $E_{F,e}$ is the Fermi energy of the electron pocket 
(Supplementary Materials Note 2 \cite{SM}). This leads to
\begin{equation}
  \nu \approx S_e \mu_e \propto T \, \frac{\sigma_e}{\sigma_h} \, \frac{\mu_e}{E_{F,e}}.
  \label{eq:nu_approx}
\end{equation}

The key insight is that the small conductivity weight of the electron pocket is 
compensated by its small Fermi energy. From the two-band parameters we find 
that the conductivity ratio between the two pockets scales with the
ratio of their Fermi energies, i.e.\ 
$\sigma_e/\sigma_h \sim E_{F,e}/E_{F,h}$ 
(Supplementary Materials Note 2 \cite{SM}). Substituting this relation into 
Eq.~(\ref{eq:nu_approx}) transfers the energy scale from $E_{F,e}$ to $E_{F,h}$, 
giving $\nu \propto T \, \mu_e / E_{F,h}$.

We now connect $E_{F,h}$ to the effective total Fermi energy $E_{F,tot}$. As discussed in Supplementary Materials Note 4 \cite{SM}, $E_{F,tot}$ is 
derived from the total carrier density $n_{tot}$ and the electronic 
specific-heat coefficient $\gamma$ via 
$T_F = \frac{\pi^2 n_{tot} k_B}{2\gamma}$ and 
$E_{F,tot} = k_B T_F$. Because $n_{tot}$ is completely dominated 
by the dense hole pocket, $E_{F,h}$ is equivalent to 
$E_{F,tot}$. Applying this second mapping, 
$\sigma_e/\sigma_h \sim E_{F,e}/E_{F,tot}$, and we obtain the final 
scaling
\begin{equation}
  \frac{\nu}{T} \approx \frac{\pi^2}{3}\frac{k_B^2}{e} \frac{\mu_e}{E_{F,tot}}.
  \label{eq:scaling}
\end{equation}
This mirrors the single-band Mott scaling $\nu/T\propto\mu/E_F$ 
\cite{behnia2009nernst}, but with $\mu$ taken as the electron mobility and the energy scale as the total Fermi energy.

This physical picture is borne out by the scaling analysis in Fig.~\ref{3}\textbf{e}. We test the scaling relation using two relevant energy scales: the electron-pocket Fermi energy $E_{F,e}$ and the effective total Fermi-energy scale $E_{F,tot}$. The data obtained with $E_{F,e}$ clearly deviate from the dashed line, whereas those plotted with $E_{F,tot}$ collapse onto the expected scaling. Here, $E_{F,h}$ is estimated to be $\sim $0.6~eV from the electronic specific-heat coefficient $\gamma$ (Supplementary Materials Note 4 \cite{SM}) \cite{EuAuBiJPS}. The large Nernst response therefore does not require invoking dominant Berry-curvature contributions \cite{BERRY,li2017,ding2019,sakai2018giant,pan2022giant}, instead, it reflects the compensation between the conductivity contrast and the Fermi-energy hierarchy intrinsic to this extreme two-pocket system.

In summary, we have revealed a clear multiband transport hierarchy in the polar semimetal EuAuBi. The dense hole pocket dominates the longitudinal resistivity, the Hall effect crosses over from electron to hole dominance with increasing field, and the Seebeck coefficient reflects a competition between the two pockets that evolves with temperature. Most notably, the magnetic-field dependence of the Nernst effect is governed by the mobility of the small pocket, while its amplitude is suppressed by the “short-circuit” effect of the large pocket. Consequently, the overall Nernst coefficient follows the scaling relation $\nu/T \propto \mu_e/E_{F,tot}$. This scaling arises from a compensation between the electron-to-hole conductivity ratio and the Fermi energy ratio, demonstrating that the transverse thermoelectric response can be dominated by a dilute, high-entropy pocket through conventional mechanisms. This understanding not only advances the fundamental transport physics of complex Fermi surfaces but also provides guidance for future studies of low-field transverse thermoelectric effects \cite{hu2025multipocket,LAGREsynergizingKMgBi}.

Weimin Quan and Xiaodong Guo contributed equally to this work. This work was supported by The National Key Research and Development Program of China (Grant No. 2024YFA1611200, 2023YFA1609600 and 2022YFA1403500), the National Science Foundation of China (Grant No. 12304065, 51821005, 12004123, 51861135104 and  11574097), the Fundamental Research Funds for the Central Universities (Grant No. 2019kfyXMBZ071), and the Hubei Provincial Natural Science Foundation ‌(2025AFA072).

\section{Data availability}
The data that support the findings of this study are available from the corresponding author upon reasonable request.

\noindent
* \verb|lixiaokang@hust.edu.cn|\\
* \verb|zengwei.zhu@hust.edu.cn|\\
\bibliography{main}

\clearpage
% Add 'S' to the numbering inside the supplement
\renewcommand{\thesection}{S\arabic{section}}
\renewcommand{\thetable}{S\arabic{table}}
\renewcommand{\thefigure}{S\arabic{figure}}
\renewcommand{\theequation}{S\arabic{equation}}
\setcounter{section}{0}
\setcounter{figure}{0}
\setcounter{table}{0}
\setcounter{equation}{0}

\begin{center}
{\large\bf Supplementary Materials for ``Multiband transport hierarchy and large Nernst effect in EuAuBi: Establishing a Nernst scaling for asymmetric multiband systems''}\\
\end{center}

\subsection*{Note 1. Materials and Methods}

Single crystals of EuAuBi with typical dimensions of ($1.3~\mathrm{mm} \times 0.8~\mathrm{mm} \times 0.6~\mathrm{mm}$) were grown using a Bi self-flux method. High-purity Eu ingots (99.9$\%$), Au ingots (99.99$\%$), and Bi ingots (99.999$\%$) were mixed in a molar ratio of Eu:Au:Bi = 1:1:10 and loaded into an alumina crucible in an argon-filled glove box \cite{EuAuBiJPS}. The crucible was then sealed in an evacuated quartz tube, heated to (1100~$^\circ\mathrm{C}$), and held there for 10 h. It was subsequently cooled slowly to (400~$^\circ\mathrm{C}$) at a rate of ($\sim 4~^\circ\mathrm{C}/\mathrm{h}$), at which point the excess Bi flux was removed by centrifugation. 

\begin{figure}[ht]
\setlength{\abovecaptionskip}{-3pt}
\setlength{\belowcaptionskip}{-3pt}
\centering
\includegraphics[width=1\linewidth]{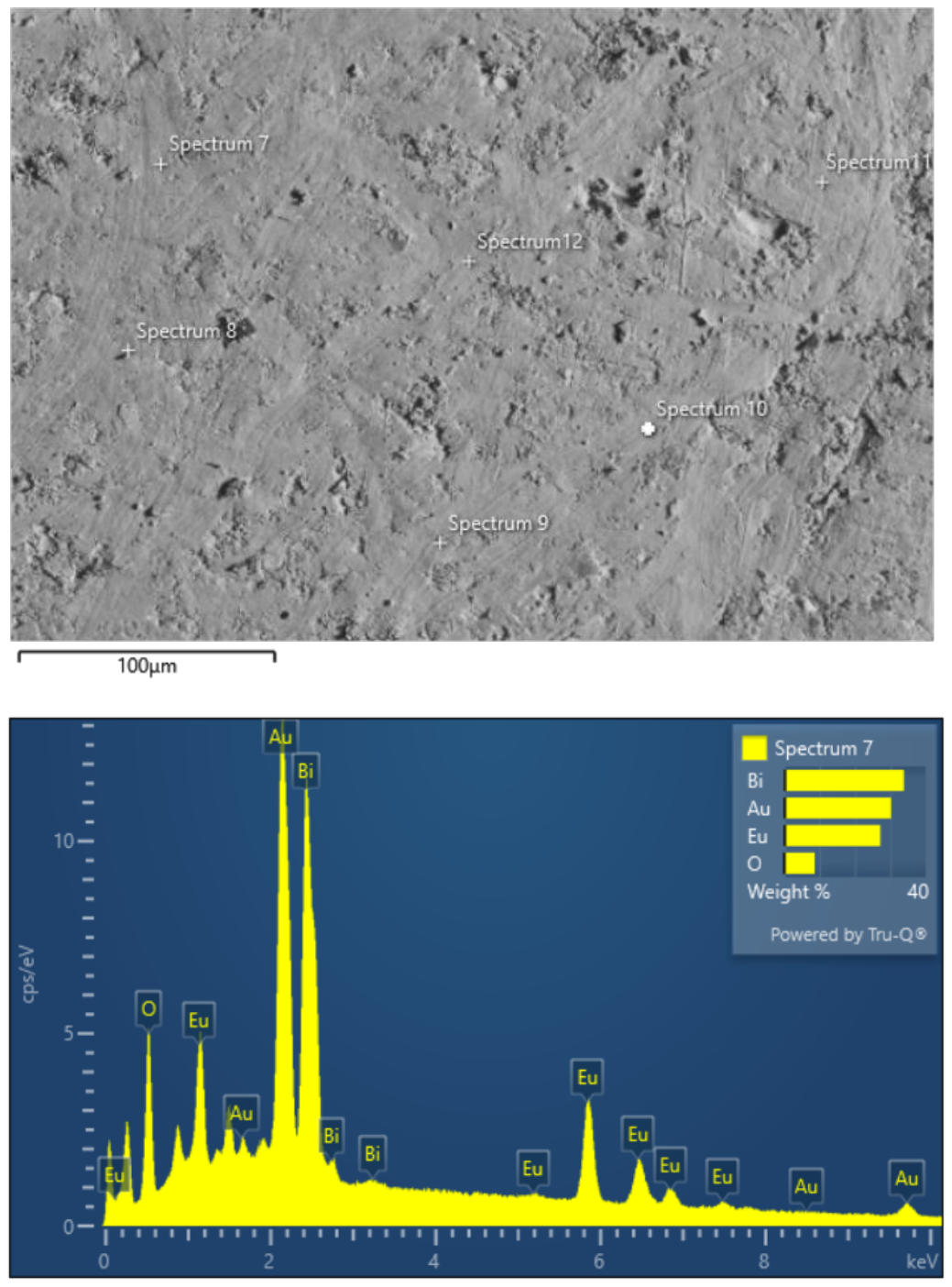}
\caption{\textbf{SEM images and EDS spectrum of the sample.}
(a) Scanning electron microscopy (SEM) images of the sample discussed in the main text.
(b) Energy-dispersive X-ray spectroscopy (EDS) spectrum extracted from the elemental mapping data of the same sample.
}
\label{S1}
\end{figure}

\begin{figure*}[ht]
\centering
\includegraphics[width=1\linewidth]{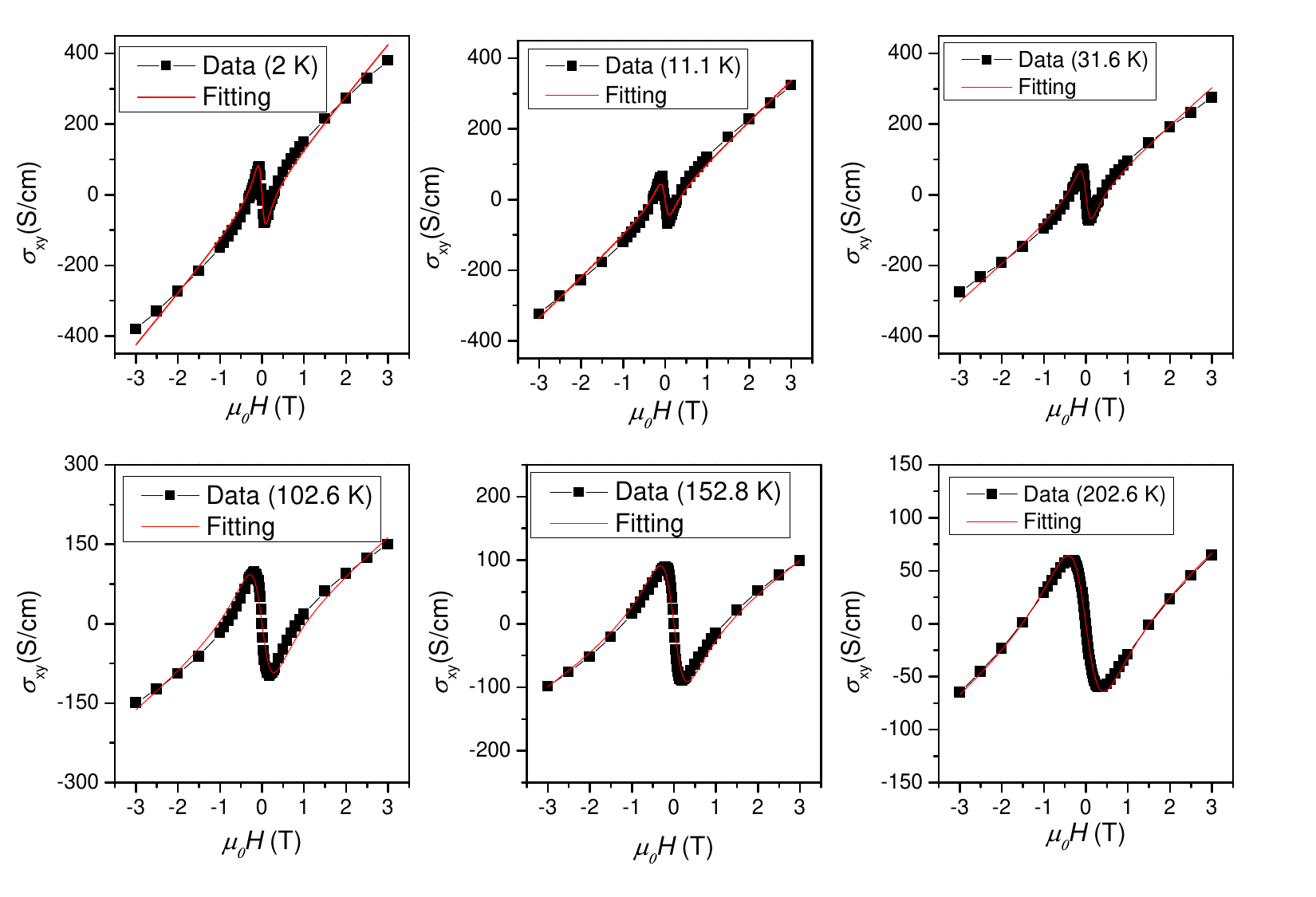}
\caption{\textbf{Two-band fitting of the Hall signals.} The Hall conductivity $\sigma_{xy}$ measured at different temperatures is well fitted by the two-band expression: $\sigma_{xy}(B) = \frac{n_e e \mu_e^2 B}{1 + (\mu_e B)^2} - \frac{n_h e \mu_h^2 B}{1 + (\mu_h B)^2}$.
}
\label{S2}
\end{figure*}

The samples discussed in the main text were also examined by energy dispersive spectroscopy (EDS) to determine their compositions. The results are listed in Fig. \ref{S1} and Tab. \ref{S2}. 

\begin{table*}[ht]
\centering
\renewcommand{\arraystretch}{1.15}
\caption{Statistics of elemental ratios in EuAuBi samples determined by EDS}
\label{tab:eds}
\begin{tabular}{lcccc}
\toprule
Statistics & O(Atomic$\%$) & Eu(Atomic$\%$) & Au(Atomic$\%$) & Bi(Atomic$\%$) \\
\midrule
Maximum    & 52.18 & 18.10 & 16.59 & 24.31 \\
Minimum    & 42.09 & 15.19 & 12.76 & 14.72 \\
Average    & 49.01 & 16.90 & 15.39 & 18.70 \\
Standard Deviation & 3.25 & 1.01 & 1.47 & 3.82 \\
\bottomrule
\end{tabular}
\end{table*}

Transport measurements were performed using a Physical Property Measurement System (Quantum Design PPMS-9). The electrical current was supplied by a Keithley 6221 current source, and the voltage was measured using a Keithley 2182A nanovoltmeter.

\subsection*{Note 2. Derivation of the two-band Nernst coefficient and the revised scaling for asymmetric multiband systems}

In a multiband system, the Nernst effect is rigorously described by the thermoelectric conductivity tensor $\alpha_{ij}$ (also called Peltier conductivity), which is additive for independent carrier pockets. The measured Nernst signal $S_{xy}$ is obtained from the resistivity and thermoelectric conductivity tensors via $\mathbf{E} = \rho \cdot \mathbf{J} + S \cdot \nabla T$ with $\mathbf{J}=0$, giving
\begin{equation}
S_{xy} = \frac{\sigma_{xx}\alpha_{xy} - \sigma_{xy}\alpha_{xx}}{\sigma_{xx}^2 + \sigma_{xy}^2}.
\label{S:eq1}
\end{equation}
Here $\sigma_{ij}$ and $\alpha_{ij}$ are the sums of the contributions from the electron ($e$) and hole ($h$) pockets.

We consider a magnetic field $B$ along $z$ and a temperature gradient along $x$. For each pocket $i$, the conductivity tensor is
\begin{equation}
\sigma_{xx}^i = \frac{\sigma_i}{1+(\mu_i B)^2},\quad \sigma_{xy}^i = \pm \frac{\sigma_i \mu_i B}{1+(\mu_i B)^2},
\label{S:eq2}
\end{equation}
where $\sigma_i = n_i e \mu_i$ is the zero-field conductivity, and $+$($-$) refers to holes (electrons). The thermoelectric conductivity has the same field dependence:
\begin{equation}
\alpha_{xx}^i = \frac{\alpha_i^0}{1+(\mu_i B)^2},\quad \alpha_{xy}^i = \pm \frac{\alpha_i^0 \mu_i B}{1+(\mu_i B)^2},
\label{S:eq3}
\end{equation}
with $\alpha_i^0$ the zero-field thermoelectric conductivity of pocket $i$. From the Mott relation,
\begin{equation}
\alpha_i^0 = -\frac{\pi^2}{3e} k_B^2 T \, \sigma_i',
\label{S:eq4}
\end{equation}
where $\sigma_i' = d\sigma_i/dE$ is on the order of $\sigma_i / E_{F,i}$ for each pocket.

In the low-field limit ($\mu_i B \ll 1$), we expand all quantities to linear order in $B$. The longitudinal and Hall conductivities become
\begin{align}
\sigma_{xx} &\approx \sigma_e + \sigma_h, \\
\sigma_{xy} &\approx (\sigma_h \mu_h - \sigma_e \mu_e) B.
\label{S:eq5}
\end{align}
Similarly,
\begin{align}
\alpha_{xx} &\approx \alpha_e^0 + \alpha_h^0, \\
\alpha_{xy} &\approx (\alpha_h^0 \mu_h - \alpha_e^0 \mu_e) B.
\label{S:eq6}
\end{align}
Inserting these into Eq.~(\ref{S:eq1}) and keeping only terms up to $O(B)$, the Nernst coefficient $\nu = S_{xy}/B$ is
\begin{equation}
\nu = \frac{ (\sigma_e+\sigma_h)(\alpha_h^0\mu_h - \alpha_e^0\mu_e) - (\alpha_e^0+\alpha_h^0)(\sigma_h\mu_h - \sigma_e\mu_e) }{ (\sigma_e+\sigma_h)^2 }.
\label{S:eq7}
\end{equation}
After algebraic simplification, this reduces to the compact form
\begin{equation}
\nu = \frac{\pi^2 k_B^2 T}{3e} \cdot \frac{ \sigma_e\sigma_h }{ (\sigma_e+\sigma_h)^2 }\, \bigl( \gamma_e + \gamma_h \bigr) \bigl( \mu_e + \mu_h \bigr) ,
\label{S:eq8}
\end{equation}
where we have introduced $\gamma_i = \sigma_i'/\sigma_i \sim 1/E_{F,i}$. Equation (\ref{S:eq8}) is the exact low-field two-band Nernst coefficient without any assumption on the relative magnitudes of the pockets.

We now apply this to EuAuBi, where $n_h \gg n_e$ and $\mu_e \gg \mu_h$, such that $\sigma_h \gg \sigma_e$ and $\gamma_e \gg \gamma_h$. In this extreme asymmetric limit,
\begin{equation}
\frac{\sigma_e\sigma_h}{(\sigma_e+\sigma_h)^2} \approx \frac{\sigma_e}{\sigma_h},\quad
\gamma_e + \gamma_h \approx \gamma_e,\quad
\mu_e + \mu_h \approx \mu_e.
\label{S:eq9}
\end{equation}
Thus,
\begin{equation}
\nu \approx \frac{\pi^2 k_B^2 T}{3e} \,\frac{\sigma_e}{\sigma_h}\, \gamma_e\, \mu_e.
\label{S:eq10}
\end{equation}
Using $\gamma_e \approx 1/E_{F,e}$ and $\sigma_e = n_e e \mu_e$, $\sigma_h = n_h e \mu_h$, we obtain
\begin{equation}
\frac{\nu}{T} \approx \frac{\pi^2 k_B^2}{3e} \cdot \frac{\mu_e}{E_{F,e}} \cdot \frac{n_e \mu_e}{n_h \mu_h}.
\label{S:eq11}
\end{equation}
This expression demonstrates that the Nernst coefficient is governed by the electron pocket's mobility and Fermi energy, but suppressed by the conductivity ratio $\sigma_e/\sigma_h$.

Equation (\ref{S:eq11}) is the rigorous low-field result for an asymmetric two-band system. i) It differs fundamentally from the conventional single-band Mott-like scaling $\nu/T \propto \mu/E_F$, which would be recovered only in the limit $\sigma_e/\sigma_h \sim 1$ (single-band or compensated systems). ii) For generic asymmetric multiband systems, the factor $\sigma_e/\sigma_h$ is essential and cannot be neglected.

Remarkably, in EuAuBi, the conductivity ratio and the Fermi energy ratio cancel each other. The total Fermi energy $E_{F,tot}$ is dominated by the dense hole pocket and can be estimated from the total carrier density and the electronic specific heat, yielding $E_{F,tot} \approx $ $0.6$ eV. The electron pocket, arising from a band crossing near the A point, has a much smaller Fermi energy $E_{F,e} \approx 20$--$30$ meV. From the extracted carrier densities and mobilities, one finds
\begin{equation}
\frac{\sigma_e}{\sigma_h} = \frac{n_e \mu_e}{n_h \mu_h} \approx \frac{E_{F,e}}{E_{F,h}} \approx \frac{E_{F,e}}{E_{F,tot}}
\label{S:eq12}
\end{equation}
within a factor of order unity. Substituting this cancellation into Eq.~(\ref{S:eq11}) gives the revised Nernst scaling for asymmetric multiband systems:
\begin{equation}
\frac{\nu}{T} \approx \frac{\pi^2}{3}\frac{k_B^2}{e} \frac{\mu_e}{E_{F,tot}}.
\label{S:eq13}
\end{equation}

Equation (\ref{S:eq13}) is the central theoretical result of this work. It establishes that in an asymmetric multiband system where one pocket dominates the total density of states while another pocket provides the dominant thermoelectric response, the Nernst coefficient scales with the mobility of the minority pocket and the total Fermi energy, rather than the mobility and Fermi energy of any single pocket. This revised scaling extends the conventional Mott-like Nernst scaling---which is strictly valid only for single-band or compensated systems---to the broad class of asymmetric multiband systems.

The cancellation that leads to Eq.~(\ref{S:eq13}) is not a generic identity but is naturally realized in systems where a small pocket coexists with a much larger conventional pocket, because the small pocket simultaneously provides small density, small Fermi energy, and high mobility. This revised scaling firmly establishes the semiclassical multiband origin of the large Nernst effect in EuAuBi.

\begin{figure*}[ht]
\centering
\includegraphics[width=1\linewidth]{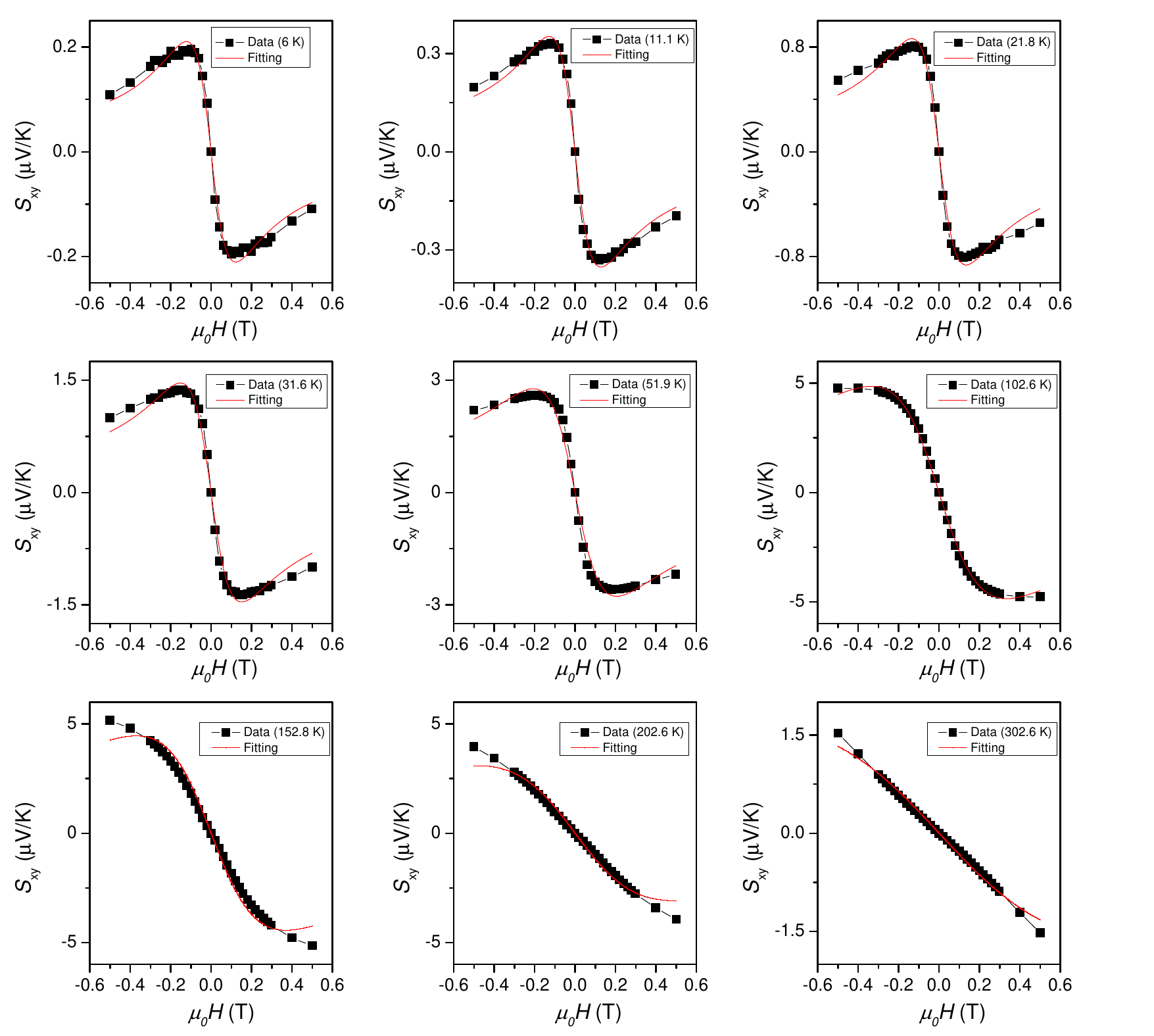}
\caption{\textbf{Single-band fitting of the low-field Nernst effect.} The Nernst signal $S_{xy}$ measured at different temperatures is well fitted below 1 T by the single-component expression $S_{xy}=\frac{S_e\mu_e B}{1+(\mu_e B)^2}$. Remarkably, the mobility $\mu_e$ is fixed to the value obtained from the Hall analysis.}
\label{S3}
\end{figure*}

\subsection*{Note 3. Fitting of the transverse Hall and Nernst signals}

The Hall conductivity at different temperatures was fitted using the two-band model (Eq. 1), and the results are summarized in Fig.~\ref{S2}.

The low-field Nernst signal $S_{xy}$ was analyzed using the single-component form
\[
S_{xy}=\frac{S_e\mu_e B}{1+(\mu_e B)^2},
\]
where $S_e$ is the effective thermopower of the electron pocket and $\mu_e$ is the corresponding mobility. In the fitting, $\mu_e$ was fixed to the value obtained from the Hall analysis rather than treated as a free parameter. Even with this constraint, the calculated curves reproduce the experimental $S_{xy}(B)$ data below $\sim 1$~T over a range of temperatures, as shown in Fig.~\ref{S3}. This consistency indicates that the low-field Nernst response originates from the same high-mobility electron pocket responsible for the low-field Hall response.

\subsection*{Note 4. Estimation of $E_{F, tot}$ from the electronic specific heat}

In Fig. 3\textbf{e}, the effective total Fermi-energy scale of EuAuBi is estimated from the electronic specific-heat coefficient $\gamma$ and the total carrier density. For a degenerate Fermi liquid, the Fermi temperature can be written as $T_F
=\frac{\pi^2 n_{\mathrm{tot}} k_B}{2\gamma}$, where $n_{\mathrm{tot}}=n_e+n_h$ is the total carrier density and $\gamma$ is the electronic specific-heat coefficient. Using $\gamma \sim 0.51~\mathrm{mJmol^{-1}K^{-2}}$ reported in Ref.~\cite{EuAuBiJPS}, we obtain the corresponding effective Fermi-energy scale $E_{F,\mathrm{tot}} = k_B T_F \approx$ 0.6 eV.

\subsection*{Note 5. The electronic structures of EuAuBi by DFT calculations}

First-principles calculations were performed within density functional theory (DFT) using the Vienna \textit{ab initio} Simulation Package (VASP) \cite{57,58}. The projector augmented-wave (PAW) method was used to describe the electron-ion interaction \cite{59}. The kinetic-energy cutoff for the plane-wave basis was set to 400~eV, and the Brillouin-zone integration was performed using a $\Gamma$-centered $9 \times 9 \times 5$ $k$-point mesh.

The exchange-correlation potential was treated within the Perdew--Burke--Ernzerhof (PBE) generalized-gradient approximation~\cite{60}. 
The Heyd--Scuseria--Ernzerhof (HSE) hybrid functional with a mixing parameter of 0.25 was further employed to obtain a more accurate description of the electronic structure \cite{61}. Spin-orbit coupling (SOC) was included in all relevant calculations.

\end{document}